# Tailoring the Electronic and Magnetic Properties of Peculiar triplet-ground-state Polybenzoid "Triangulene"

Vaishali Sharma[1], Som Narayan[1], Shweta D Dabhi[2] and Prafulla K Jha[1,#]

[1]Department of Physics, Faculty of Science, The M. S. University of Baroda, Vadodara-390 002, India
[2]Department of Physics, MK Bhavnagar University, Bhavnagar-364 001, India

#Email: prafullaj@yahoo.com

## Abstract

In the present work we have studied the structural and electronic properties of recently synthesized elusive free standing triangulene using density functional theory. Triangulene, which is a type of graphene quantum dot, is a molecule with an even number of electrons and atoms but the structure of molecule is such that it is impossible to pair all these electrons. The spins of these two unpaired electrons have two possible orientations: triplet (ferromagnetic) and singlet (antiferromagnetic) state. From the first principles study of free standing triangulene, we found energetically, by koopman's theory of global reactivity descriptors and frequency calculations that triplet (ferromagnetic) is more stable than singlet (antiferromagnetic) which is in good agreement with the previous results. These elementary studies are technologically compatible as open shell graphene quantum dots could be useful in spintronic and magnetic carbon materials. Further we have also studied the influence of magnetic elements Fe, Co, Ni and Cu on triangulene for their applications in spintronics. Our results suggest that the transitional metal (TM) doped graphene quantum dot is interesting for information readout devices where the Tm-ion spin states can be used to store information.

# 1. Introduction

Over the past decade, among all the nanomaterials, graphene[1] takes a special place because of its sumptuous properties which results in wide range of applications[2] such as organic semiconductors, chemical sensors, biological engineering, composite materials, energy storage devices, photovoltaic cells, in spintronic and nonlinear optics[2-3]. However, the major problem with the graphene is its direct electronic applications and due to the absence of electronic band gap in it. The band gap in graphene can be opened by controlling the size, such as a nano-ribbon strip and by introduction of other hetero atoms and defects into the graphene moiety[4]. The graphene quantum dots (GQDs) that consist of nanometer-scaled graphene particles with $sp^2$ carbon bonds show great potential in this regard. At this stage most of the applications for GQDs are focused on photoluminescence (PL) related fields energy and display[5-6]. The graphene quantum dots regarded as small piece of graphene are a kind of zero dimensional (0D) material with characteristics derived from both graphene and carbon dots (CDs)[7-8.] The GQDs are superior in terms of high photostability, biocompatibility and low toxicity than organic dyes and semiconducting quantum dots (SQDs) due to quantum confinement and edge effects[7-9].

The tiny fragments of graphene sheet results into the nanographene with various polycyclic aromatic hydrocarbon (PAH) units. Therefore the properties of PAHs are closely related to nanographenes[10-11]. These PAHs are made up of fused aromatic rings which lead to the features with almost infinite possibilities that can lead to a rich variety of compounds. There are types of PAHs which acquires a high-spin open-shell radical character in their ground state[11]. For example, phenalenyl[12] contains an odd number of carbon atoms with an odd number of π electrons which makes it a radical in its neutral ground state. The extension of benzene rings in a triangular form can lead to several π-conjugated phenalenyl derivatives such as triangulene (also known as clar's hydrocarbon). Triangulene is the smallest triplet-ground-state polybenzoid, exists as a diradical containing an even number of carbons (22, in fused benzene rings), has been an obscure molecule ever since it was first hypothesized[13]. It is impossible to draw kekulè structure for the whole molecule, two unpaired valence electrons results in every attempt. It is a system containing even number of π-electrons although the topology of the system means that it is impossible to write a resonance structure in which each π-electron is paired with one on a neighbouring carbon[14]. Due to its extreme reactivity[13] synthesis and the characterization of unsubstituted triangulene has not been achieved. Recently, triangulene by manipulating a precursor molecule (from a mixture of dihydrobenzo[cd,mn]pyrene isomers) has been successfully created on Cu(111), NaCl (100) and Xe 111)[15]. They showed that STM, DFT and AFM calculations confirmed that the triangulene keeps its free molecule properties on the surface and exists in threefold symmetric molecular structure[15].The expected, aligned spins- the quantum-mechanical property that gives electrons a magnetic orientation due to two unpaired electrons, has inquested triangulene's magnetic properties. This property could make triangulene useful in electronics, applications in quantum computing, quantum information processing and spintronics[16-17]. Though the supportive DFT calculations clearly brings out the free molecule properties of triangulene deposited on three surfaces NaCl (100), Cu (111) and Xe (111), a detail investigation on the physical properties of free triangulene is still awaited. Pavlicek et al.[15] have also emphasized for a thorough investigation of the spin ground state and excited states. Motivated with this fact we conduct a systematic first principles study of graphene quantum dots, triangulene using density functional theory.

Furthermore, the magnetism of triangulene and its derivative, because of their open shell (unpaired electron) character has aroused continuous interest for potential applications in spintronics[16]. Recently, there has been great interest in spin-based electronics (spintronics), in which it uses the spin degree of freedom of electrons instead of charges as the carrier of information. The strong diamagnetic character of graphene induces magnetic ordering which is of great importance. Up to the date, several researchers theoretically and experimentally found the change in magnetic moment of graphene by vacancy defects[18-20], heavy and light adatoms[21-22], transition metals[23] and also by the edge effect[24]. The diamagnetic behaviour of graphene overcomes as we go to low dimensional graphene quantum dots[25]. Graphene quantum dots in which charge carriers are confined provides full control over an individual electron, which in turn plays an important role in investigating the behaviour of spins[26]. To the best of our knowledge the nature of magnetic interactions in triangulene has not been explored yet. Therefore, to tune the magnetic moment of GQDs, we have further studied the influence of iron (Fe), cobalt (Co), nickel (Ni), and copper (Cu) on triangulene.

The objective of the present study is three fold (i) to perform the study of electronic and vibrational properties of free triangulene in ground state, (ii) perform studies on the effect of four transition metals (Fe, Co, Ni and Cu) on pristine triangulene's electronic and magnetic properties. This will help to understand the magnetic states in triangulene where the edges are significant and resulting into high edge to area ratio, (iii) finally to extend the calculations to determine the different global reactivity descriptors such as polarizability, global hardness, electrophilicity and additional electronic charge of triangulene.

# 2. Computational methodology

The full geometry optimization, electronic density of states (DOS) and vibrational spectra presented here in has been carried out using closed-shell Becke–Lee–Yang–Parr hybrid exchange–correlation three-parameter functional (B3LYP)[27] with the 6-31G basis set under the frame work of density functional theory (DFT) based Gaussian 09 suite of program[28]. All structures have been fully relaxed in the absence of external fields. The vibrational frequencies and Raman intensities of triangulene were calculated. GaussView[29] program has been used for the pictorial visualization, atomic orbital compositions of molecular orbitals, checking of calculated data and to get the graphical presentation of isoelectronic molecular electrostatic potential surfaces (EPS), Raman, IR spectra. The electronic parameters, such as lowest unoccupied molecular orbital (LUMO) energy and highest occupied molecular orbital (HOMO) energy and the band gap energy ($E_g = \varepsilon_{LUMO} - \varepsilon_{HOMO}$), were obtained through theoretical calculations. The chemical reactivity of a triangulene is described by global electronic reactivity descriptors. According to Koopman's theorem[27], the different global reactivity descriptors, i.e., electronegativity (χ), chemical potential(μ), global hardness (η), global electrophilicity index (ω) and global softness (S) are evaluated by means of the energies of frontier molecular orbitals $\varepsilon_{LUMO}$, $\varepsilon_{HOMO}$ and are given by[30-31];

$$\chi = -\frac{1}{2}\,(\epsilon_{HOMO} + \epsilon_{LUMO}) \qquad (1)$$

$$\mu = -\chi = \frac{1}{2}\,(\epsilon_{HOMO} + \epsilon_{LUMO}) \qquad (2)$$

$$\eta = \frac{1}{2}\,(\epsilon_{LUMO} - \epsilon_{HOMO}) \qquad (3)$$

$$S = \frac{1}{2\eta} \qquad (4)$$

$$\omega = \frac{\mu^2}{2\eta} \qquad (5)$$

$$\Delta N_{max} = -\frac{\mu}{\eta} \qquad (6)$$

The energies ω, μ, χ, η, S, $\varepsilon_{HOMO}$, $\varepsilon_{LUMO}$ and energy band gap ($\varepsilon_{LUMO} - \varepsilon_{HOMO}$) for triangulene are listed in Table 2. The adsorption energy ($\Delta E_{ad}$) of the transition metals Fe, Co, Ni and Cu to the triangulene has been calculated as follows:

$$(\Delta E_{ad}) = E_{triangulene+Fe/Co/Ni/Cu} - \left(E_{triangulene} + E_{Fe/Co/Ni/Cu}\right) \qquad (7)$$

Where $E_{\mathrm{triangulene+Fe/Co/Ni/Cu}}$ are the total energies of interacting triangulene-transition metals. $E_{triangulene}$ is the total energy of triangulene while $E_{\mathrm{Fe/Co/Ni/Cu}}$ is the total energy of an individual transition metal. It should be mentioned that the adsorption energies have corrected for the basis set superposition error (BSSE) by means of the Boys–Bernardi counterpoise method to eliminate basis function overlap effects[33].

# 3. Results and discussion

The structure of an odd molecule triangulene and its isomers has been recently reported[15]. The optimized structure of triangulene for antiferromagnetic singlet and ferromagnetic triplet states obtained at the B3LYP/6-31G level is shown in Figs. 1 (a) and (b) respectively. As can be seen from the enlarged section of these figures, the two particular bond lengths are different for open shell triplet and singlet states. The bond lengths 10C-7C and 7C-3C are with same values 1.412 Å in the more stable triplet ground state than the same corresponding length 10C-7C (1.422 Å) and 7C-3C (1.40 Å) in singlet ground state. These can be attributed to the stability of the FM structure as well as confirm that the shapes and edges which depends on the bond lengths and angles in the benzene ring affect magnetism in triangulene like other GQDs[34-36]. Fig 2 presents the optimized structures of triangulene with four transition metals. In our initial configuration to perform the DFT calculation, transition metals were on top-site of 8C (numbering can be seen from Fig. 1) carbon atom in triangulene with the distance around 1.8 Å. It should be noted that after optimization the transition metals are reoriented at the outer side of triangulene, this can be attributed to the fact that the outer side of triangulene is more electropositive (discussed latter) and the transition metals are highly electronegative. It is interesting to find that Fe, Co and Cu moves away from the 8C carbon atom and migrates to the 16C carbon atom making bond with the bond distance 1.95 Å, 1.91 Å, 2.05 Å and the bond angle (17H16C35Fe/Co/Cu) of 130°, 100° and 125° respectively. However, Ni moves towards the unpaired electron (7C) making bond with bond distance 1.94 Å and bond angle (33H7C35Ni) 127.5°. Table 1 shows the calculated energy and dipole moment of free standing triangulene. Both structures of triangulene reveal that the energetically ferromagnetic triplet state is more stable than the antiferromagnetic singlet state in agreement with the previous work[15]. Further we studied the alpha and beta phase of FM triangulene as it is the more stable one energetically. Alpha and beta ferromagnetic are the spin up and spin-down orbitals. As triangulene consist of two unpaired electrons, electrons can have two spin alpha (+1/2) and beta (-1/2). Results in Table 1 shows that the adsorption of Fe, Co, Ni and Cu on triangulene has increased the dipole moment value as compared with pristine triangulene. The dipole moment in a molecule is a substantial property which provides information regarding the change in electronic distribution upon excitation. The total energy value is increased with all various adsorption of Fe, Co, Ni and Co, which leads to make the triangulene structure more stable. The calculated molecular dipole moment of ferromagnetic (FM) and antiferromagnetic (AFM) states are 0.0002 and 0.0325 Debye respectively which shows that the AFM triangulene has almost 0.0323 Debye higher value of dipole moment than FM triangulene. This shows that the molecular displacement has more significant contribution from the lone pair electron than the electron shared out more diffusely in covalent bond. Further the higher dipole moment for AFM triangulene indicates its higher activity and lesser stability as compared to FM triangulene, a fact that is also confirmed from the energy presented in Table 1. There is no net dipole moment because individual bond dipole moments completely cancel out each other in the case of highly symmetrical molecular geometries. An electron pair has no dipole moment as the two electrons have opposite spins resulting their magnetic dipole fields in opposite directions and cancel. But an unpaired electron has a magnetic dipole and interacts with a magnetic field. Therefore, due to the structure of triangulene, it always has two unpaired electrons in which the bond dipole moments cannot cancel one another. Consequently, molecules with these geometries always have a nonzero dipole moment[37].

As shown in Table 3., the calculated adsorption energy of Fe, Co, Ni and Cu on triangulene are -11.7 eV, -9.77 eV, -10.53 eV and -17.58 eV respectively, which are stronger than they adsorbed on graphene, bilayer graphene, CNT and $MOS_2$[21,38-40]. From the definition of adsorption energy, the negative value of $\Delta E_{ad,}$ the stronger the interaction between transition metals and triangulene. Fig. 2 depicts that after optimization, transition metals (Fe, Co, Ni and Cu) chemically bound to the surface making covalent bond leading to the chemisorption in triangulene. The adsorption nature of transition metal over triangulene are in the order of Cu > Fe > Ni > Co. Moving on we focus on the magnetic moments induced by transition metals in triangulene. DFT calculations give the atomic magnetization M in case of stable spin states. Table 3 shows the change in magnetic moment of triangulene with Fe, Co, Ni and Cu. The atomic magnetization of FM phase of triangulene is $4.23\mu_{B,}$ which is greater due to their unpaired electron. Further the magnetic moment is modified with the adsorption of Fe, Co, Ni and Cu with the values 4.8 $\mu_{B,}$ 4.82 $\mu_B$, 4.85$\mu_B$ and 4.857 $\mu_B$ respectively. The magnetic moment increases by the value of around 0.62 $\mu_B$ in all systems considered. Triangulene gives large magnetization with transition metals, greater than others[21,40-42].

The calculated Raman spectra are presented in Fig. 3, where the calculated intensity is plotted against harmonic vibrational wavenumbers. On the basis of our calculations AFM and FM both triangulene have no imaginary frequencies. Thus the stability of the optimized geometries is confirmed by frequency calculations which give positive values for all obtained frequencies. For both AFM and FM phases of triangulene, the most intense peaks are G and D, appearing around 1596 $cm^{-1}$ and 1350 $cm^{-1}$, respectively. The G peak is due to the C-C bond stretching, i.e. the first order Raman-allowed $E_{2g}$ phonon at the Brillouin zone centre. D peak is found in both triangulene states due to the finite crystalline size where the edges can be seen as defects. Also a localized mode is found to appear at ≈ 3220 $cm^{-1}$ in triangulene, which is the typical vibrational mode of the C-H bond[43]. The vibrational analysis of free standing triangulene with their Raman and IR are presented in Table 4. In the frequency range around 3200 $cm^{-1}$, we notice the major Raman spectra due to C-H stretching vibrations in FM triangulene. In the range between 1000-3230 $cm^{-1}$ some weak bands are found due to C-C and C-H stretching vibrations. In this frequency range, the infrared spectra are 72 $cm^{-1}$ and 130$cm^{-1}$ due to C-H out-of-plane bending vibrations of triangulene (FM). However, in triangulene (AFM), we found major intense Raman peaks between 50-500 $cm^{-1}$ due to C=C, C-C and in-plane stretching that leads to the ring deformation. In the IR spectra bands at 107 and 91 $cm^{-1}$ are found to be associated with C-H, C=C out-of-plane bending and in-plane stretching vibrations. All the values are depicted in Table 4. Also Fig. 8 shows the various molecular vibrations with Raman and IR spectroscopy.

Fig. 4(a) shows the electronic density of states (DOS) of AFM, alpha (up-spin) and beta (down-spin) for graphene quantum dots triangulene. The DOS clearly depicts the 0D structure (quantum dot) as all available electronic states exist only at discrete energies and can be represented by delta function. Fig. 4(b) presents the density of states of triangulene on adsorption of transition metals Fe, Co, Ni and Cu respectively. The dashed line in both DOS graphs represents the HOMO energy. We can clearly see the shift in HOMO energies from the graph. The pattern of the highest occupied molecular orbital (HOMO) and the lowest unoccupied molecular orbital (LUMO) for the triangulene in AFM, alpha (up-spin) and beta (down-spin) FM phases are shown in Fig. 5 and their energy gaps are found to be 0.738 eV, 4.234 eV and 3.968 eV respectively. Ideally both alpha and beta FM should have same value of HOMO-LUMO but as the two unpaired electrons are apart from each other it contributes differently in both alpha and beta phases. It is observed from this figure that the FM state has higher energy gap. The HOMO-LUMO gap can be used to describe the chemical stability and electrical transport properties of triangulene. A large and small gap shows the high and low stability respectively. Therefore, it is consistent with above discussion that the triangulene in FM state is more stable. Furthermore, a molecule with a small HOMO-LUMO gap is more polarizable (reactive) indicating more reactivity or instability for AFM state triangulene[44-45]. Fig. 6 presents HOMO-LUMO of triangulene with Fe, Co, Ni and Cu with the values 2.7 eV, 3.5 eV, 2.8 eV and 1.28 eV respectively. There is a fluctuating flow in these energy gaps as from going Fe to Co the energy gap increases and decreases in moving from Co to Cu. This is due to the fact that addition of electrons leads to a reduction in energy gap. So the removal of electrons gives the abnormal behaviour in HOMO-LUMO gaps. Applying the koopman's theorem, global hardness ($\eta$), softness (S) values, electrophilicity index ($\omega$), chemical potential ($\mu$), additional electronic charge ($\Delta$N) from the neighbouring atoms are calculated and presented in Table 2. According to the maximum hardness principle, the most stable structure has maximum hardness. Therefore, the AFM triangulene with minimum $\eta$ value will be comparatively less stable than FM triangulene. The global hardness($\eta$) is highest for beta FM. Global electrophilicity index ($\omega$) measures the stabilization in energy when the system acquires an additional electronic charge from the neighbours. The path of the charge transfer as an electrophile (is a chemical reagent efficient of accepting electrons from surroundings) is entirely determined by electronic chemical potential of the molecule. Hence its electronic chemical potential must be negative and its energy must be lower upon accepting electronic charge. The calculated high value of $\omega$ shows that the ferromagnetic behaves as a strong electrophile than antiferromagnetic phase of triangulene. The molecular electrostatic potential (MESP) which can be calculated from theoretically derived electron density is an important quantity to understand the interactive behaviour and properties of a molecule[46-47]. The most negative valued points in the MESP topography $V_{min}$ is widely used to gauge the electron donating properties of a molecule, while MESP at a nucleus $V_n$ measures the interactive behaviour of a molecule with respect to a particular atom towards the electron rich/electron deficient site of another molecule. MESP is very useful for predicting the site of chemical reactivity, dipole moment, electronegativity and partial charges of the molecule. The molecular electrostatic potential (MESP) at a point *r* in the vicinity of a molecule (in atomic units) can be expressed as,

$$V(r) = \sum \frac{Z_A}{|R_A - r|} - \int \frac{\rho(r')\, dr'}{|r' - r|} \quad (8)$$

Where $Z_A$the charge on nucleus A is positioned at $R_A$ and $\rho(r')$ is the density at any point *r*. The former in the expression shows the effect of the nuclei and latter the effect of electrons. $V(r)$ shows the net electrostatic effect produced at the point *r* by the total charge distribution (electrons + nuclei) of the molecule. MESP properties of triangulene are calculated at 6-31G level using Gaussian 09[50]. Fig. 7 presents the MESP surface o AFM and FM triangulene. The colour coding in Fig. 5 indicates the values of MESP and are automatically chosen to represent the most negative (coloured in the shades of red in the surface) and the most positive (coloured in the shades of blue) electrostatic potential also the green region represents the zero potential on that surface. Furthermore, the positive MESP resembles the repulsion of the proton by atomic nuclei in field of low density and the nuclear is incompletely guarded. The negative MESP resembles the attraction of proton by the concentrated electron density in the molecule. The MESP at different points on the electron density isosurface is presented by colouring the isosurface with contours. These potential increases in the order of red < orange < yellow < green < blue.

# 4. Conclusions

Being very evasive experimentally, triangulene has been an important graphene quantum dot for theoretical investigations. In triangulene there are two $\pi$ – electrons that cannot be paired (Parr et al. 1922). Resulting that the spins of these two electrons have two possible orientations: singlet (antiferromagnetic) and triplet (ferromagnetic). From the first principles study of free standing triangulene, among the two spin states, we found that ferromagnetic state is more stable than antiferromagnetic state. We concluded it by energetically and global reactivity descriptors obtained from Koopman's theorem. The HOMO-LUMO gap and global hardness are higher for stable FM triangulene. The positive vibrational frequencies also confirm the stability of triangulene. The discrete energies in density of states indicates the three dimensional confinement and hence is a graphene quantum dot. The large magnetization leads to their further applications in spintronic devices. The modification of magnetic moment gives better results in case of triangulene with transition metals (Fe, Co, Ni and Cu adatoms) in which the unique topology of triangulene gives rise to high-spin ground states, conceivably useful in organic spintronic devices. On the basis of our studies, if the concentration of adatoms is high, a magnetic island presenting interesting magnetic behaviour can be formed.

**Conflict of Interest**: The authors declare no competing financial interest.

# 5. Acknowledgement

Authors are thankful to SERB (SR/S2/CMP-0005/2013) and DST (DST/INT/POL/P-33/2016) for the financial support.

## Table Captions

**Table 1:** Calculated Energy and Dipole Moment for pristine triangulene and with Fe, Co, Ni and Cu.

**Table 2:** Calculated energy of HOMO-LUMO (eV), Energy gap($\varepsilon_{HOMO}$-$\varepsilon_{LUMO}$), electronegativity($\chi$), chemical potential ($\mu$), global hardness ($\eta$) and softness (S), global electrophilicity index($\omega$), additional electronic charge ($\Delta$Nmax)

**Table 3:** Calculated HOMO-LUMO energy, HOMO-LUMO gap, adsorption energy and magnetic moment of triangulene with Fe, Co, Ni and Cu.

**Table 4:** Calculated Frequency of triangulene (FM) and (AFM) with IR and Raman vibrations.

## Figure Captions

**Figure 1:** Optimized geometry of free standing triangulene (a) Antiferromagnetic (b) Ferromagnetic. The blue and red ball represents the carbon and hydrogen atoms.

**Figure 2:** Top and side view of optimized structure of triangulene with (a) Fe (b) Co (c) Ni (d) Cu. The green, fluorescent, orange, and light blue ball represents Fe, Co, Ni and Cu atoms.

**Figure 3:** Raman spectra of triangulene (a) AFM (b) FM.

**Figure 4:** (a) Density of states of triangulene (AFM and FM). (b) DOS of triangulene with the adsorption of (a) Fe (b) Co (c) Ni (d) Cu

**Figure 5:** Schematic representation of the HOMO and LUMO level of triangulene showing the change in the HOMO−LUMO gap (a) antiferromagnetic (b) alpha ferromagnetic (c) beta ferromagnetic. The dotted line indicates the higher energy level.

**Figure 6:** Schematic representation of the HOMO and LUMO level of triangulene showing the change in the HOMO−LUMO gap with (a) Fe (b) Co (c) Ni (d) Cu. The dotted line indicates the higher energy

**Figure 7: (a)** Molecular Electrostatic Potential mapped on the isodensity surface for AFM triangulene in the range from $-2.102 \times 10^{-2}$ (red) to $+2.102 \times 10^{-2}$ (blue). **(b)** Molecular Electrostatic Potential mapped on the isodensity surface for FM triangulene in the range from $-2.043 \times 10^{-2}$ (red) to $+2.043 \times 10^{-2}$ (blue).

**Figure 8:** Various molecular vibrations with Raman and IR. Figs (a-c) show the out-of-plane vibrations and Figs. (d-f) shows in-plane vibrations.

**Table 1**

| Triangulene | Energy (eV) | Dipole Moment (Debye) |
|---|---|---|
| Antiferromagnetic ( AFM ) | -23004.41 | 0.0325 |
| Ferromagnetic (FM) | -23004.72 | 0.0002 |
| Triangulene+Fe | -57369.76 | 3.266 |
| Triangulene+Co | -60611.14 | 3.355 |
| Triangulene+Ni | -64026.12 | 1.727 |
| Triangulene+Cu | -67620.42 | 2.2443 |

**Table 2**

| Triangulene | HOMO (eV) | LUMO (eV) | $E_g$ (eV) | χ (eV) | μ (eV) | η (eV) | S (1/eV) | ω (eV) | $\Delta N_{max}$ (eV) |
|---|---|---|---|---|---|---|---|---|---|
| Antiferromagnetic | -3.73 | -2.99 | 0.73 | 3.36 | -3.36 | 0.36 | -1.35 | -15.37 | -9.12 |
| Alpha Ferromagnetic | -3.74 | 0.4946 | 4.23 | 1.62 | -1.62 | 2.11 | -0.23 | -0.621 | -0.76 |
| Beta Ferromagnetic | -6.11 | -2.1496 | 3.96 | 4.13 | -4.13 | 1.98 | -0.25 | -4.305 | -2.08 |

**Table 3**

| System | $E_{HOMO}$ (eV) | $E_{LUMO}$ (eV) | $E_g$ (eV) | $\Delta E_{ad}$ (eV) | Magnetic Moment ($\mu_B$) |
|---|---|---|---|---|---|
| Triangulene + Fe | -12.8 | -10.09 | 2.7 | -11.7 | 4.8 |
| Triangulene + Co | -13.3 | -9.80 | 3.5 | -9.77 | 4.82 |
| Triangulene + Ni | -12.86 | -10.03 | 2.83 | -10.5 | 4.85 |
| Triangulene + Cu | -10.57 | -9.28 | 1.28 | -17.58 | 4.857 |

**Table 4**

| Triplet States (FM) | | Singlet States (AFM) | |
|---|---|---|---|
| **Cal. Bands IR/Raman** | **Description** | **Cal. Bands IR/Raman** | **Description** |
| $84^{R}$ | (20C=23C-21C), (27C-29C-31H), (10C=9C-11C), (4C=8C-14C), (4C=8C-9C), (14C-8C-9C), (10C-20C=23C), (4C-3C-7C), (14C=13C-12C) s stretching | $138179^{R}$ | (10C-20C=23C), 23C-21C=11C), (10C=9C-11C), (2C=3C-7C), (12C-13C-28C), (5C=16C-15C), (4C=8C-14C) a. s. stretching |
| $72^{IR}$ | (2C-26H)+(1C-18H)+ (6C-19H)+ (27C-30H)+ (29C-31H)+ (28C-32H)+(21C-22H) wag o.p | $9590^{R}$ | (10C-20C)+ (11C-21C) rocking i.p |
| $130^{IR}$ | (3C-7C-33H), (13C-12C-34H), (1C-18H), (29C-31H) wag o.p | $1531^{R}$ | (21C-22H), (20C-25H), (10C=9C-11C) twisting o.p |
| $60^{R}$ | (20C=23C-21C) a.s stretching, (15C-14C=13C) (3C-7C-10C), (1C-2C=3C) s. stretching, (20C-25H), (21C-22H) rocking i.p | $24611^{R}$ | All rings i.p rocking + ring deformation |
| $27^{R}$ | (10C=9C-11C) scissoring i.p (20C=23C-21C), (4C=8C-14C) a.s stretching, (16C=5C-4C),(1C-2C=3C) scissoring i.p +ring deform | $14^{IR}$ | (2C-26H)+(1C-18H)+ (6C-19H)+ (27C-30H)+ (29C-31H)+ (28C-32H)+(21C-22H) wag o.p |
| $18^{R}$ | (4C=8C-14C), (20C=23C-21C), (10C=9C-11C)a.s stretching, (3C-7C-10C), (13C-12C-11C)s. stretching, (6C=1C-2C), (29C=28C13C) scissoring i.p + ring deform | $2880^{R}$ | (20C=23C-21C), (23C-21C=11C), (3C-7C-10C), (11C-12C-13C) scissoring i.p |
| $25^{R}$ | (9C-8C-14C) scissoring i.p, (26H-2C), (1C-18H), (6C-19H), (29C-31H), (28C-32H), (20C-25H), (21C-22H) rocking i.p | $61^{IR}$ | (25H-20C=23C-24H), (24H-23C-21C-22H), (26H-2C-1C-18H), (30H-27C-29C-31H) wag o.p |
| $78^{R}$ | (20C-25H), 23C-24H), 21C-22H) rocking i.p, (13C-12C-11C), (10C-7C-3C), (15C-16C=5C) s. stretching, (12C-(13C=14C), (7C-3C-4C), 27C=15C-16C), (16C=5C-6C) a.s stretching + ring deformation | $107^{IR}$ | (25H-20C=23C-24H), (24H-23C-21C-22H), (26H-2C-1C-18H), (30H-27C-29C-31H) twisting o.p |
| $238^{R}$ | (20C-25H), (21C-22H) rocking i.p, (2C=3C-4C), (14C=13C-28C), (2C-1C=6C), (27C-28C=29C)a.s stretching, | $2303^{R}$ | (10C-20C-25H), (11C=21C-22H), (27C-30H), (28C-32H), (6C-19H), (2C-26H) rocking i.p, (4C=8C-14C) |
| | (4C=8C-14C)s.stretching, (5C=16C-15C) + ring deform | | scissoring i.p + ring deformation |
| 213[R] | (2C-26H), (20C-25H), (21C-22H), (6C-19H) s. stretching | 672[R] | (26H-2C), (6C-19H), (27C-30H), (28C-32H) rocking i.p, (5C=16C-15C), (10C=9C-11C) scissoring i.p |
| 81[IR] | (18H-1C-2C-26H), (31H-29C=28C-32H) a.s. stretching, (18H-1C=6C-19H), (30H-27C-29C-31H) s. stretching | 976[R] | (23C-24H), (12C-34H), (7C-33H), (16C-17H) rocking i.p, (10C=9C-11C), (4C=8C-14C), 3C-7C-10C), (11C-12C-13C) a.s. stretching |
| 1054[R] | (20C-25H), (23C-24H), (21C-22H), (2C-26H), (1C-18H), (28C-32H), (29C-31H), (6C-19H), (27C-30H) s. stretching | 91[IR] | (20C-25H), (23C-24H), (21C-22H), (2C-26H), (28C-32H), (29C-31H), (1C-18H) (6C-19H), (27C-30H) s. stretching |

All frequencies are in $cm^{-1}$. s, a.s, i.p and o.p denotes symmetry, anti-symmetry, in-plane and out-of-plane.

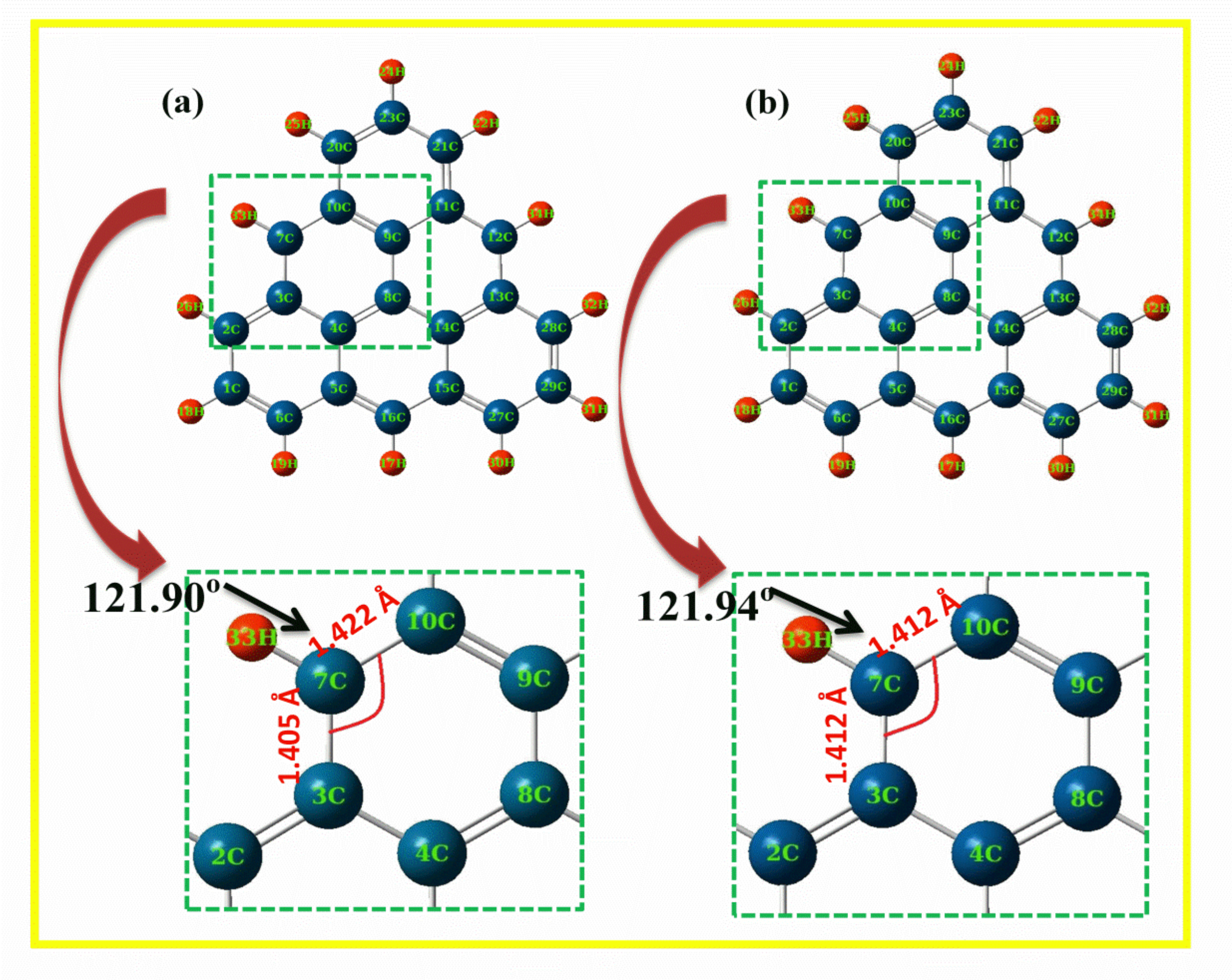

**Fig. 1**

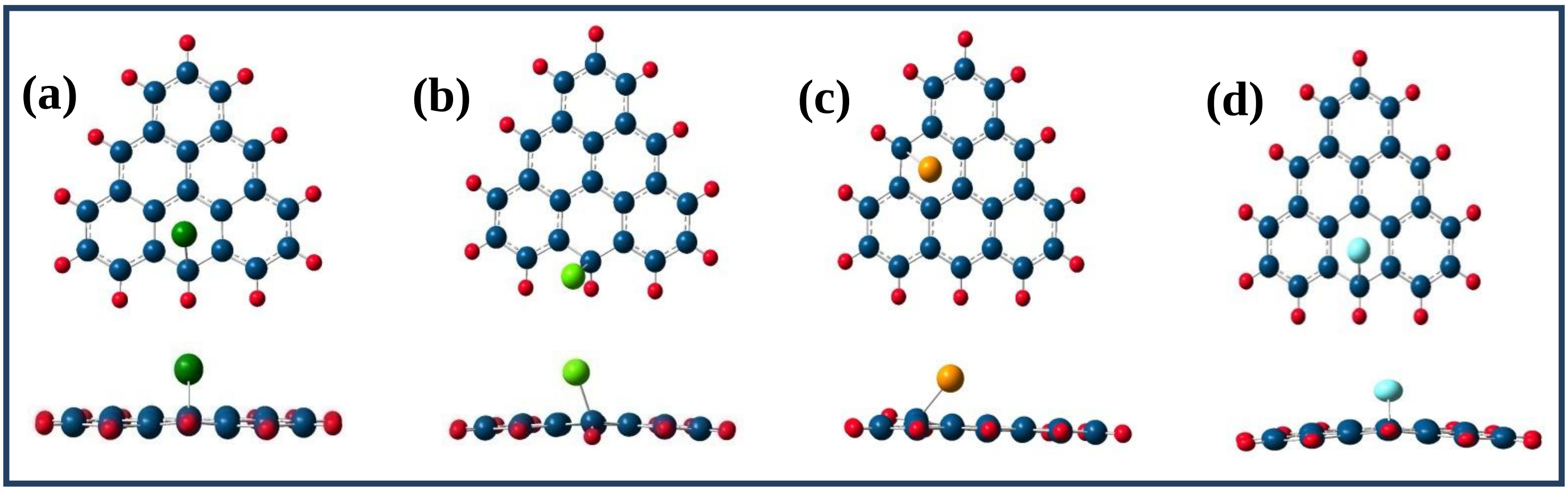
(a)
(b)
(c)
(d)

**Fig. 2**

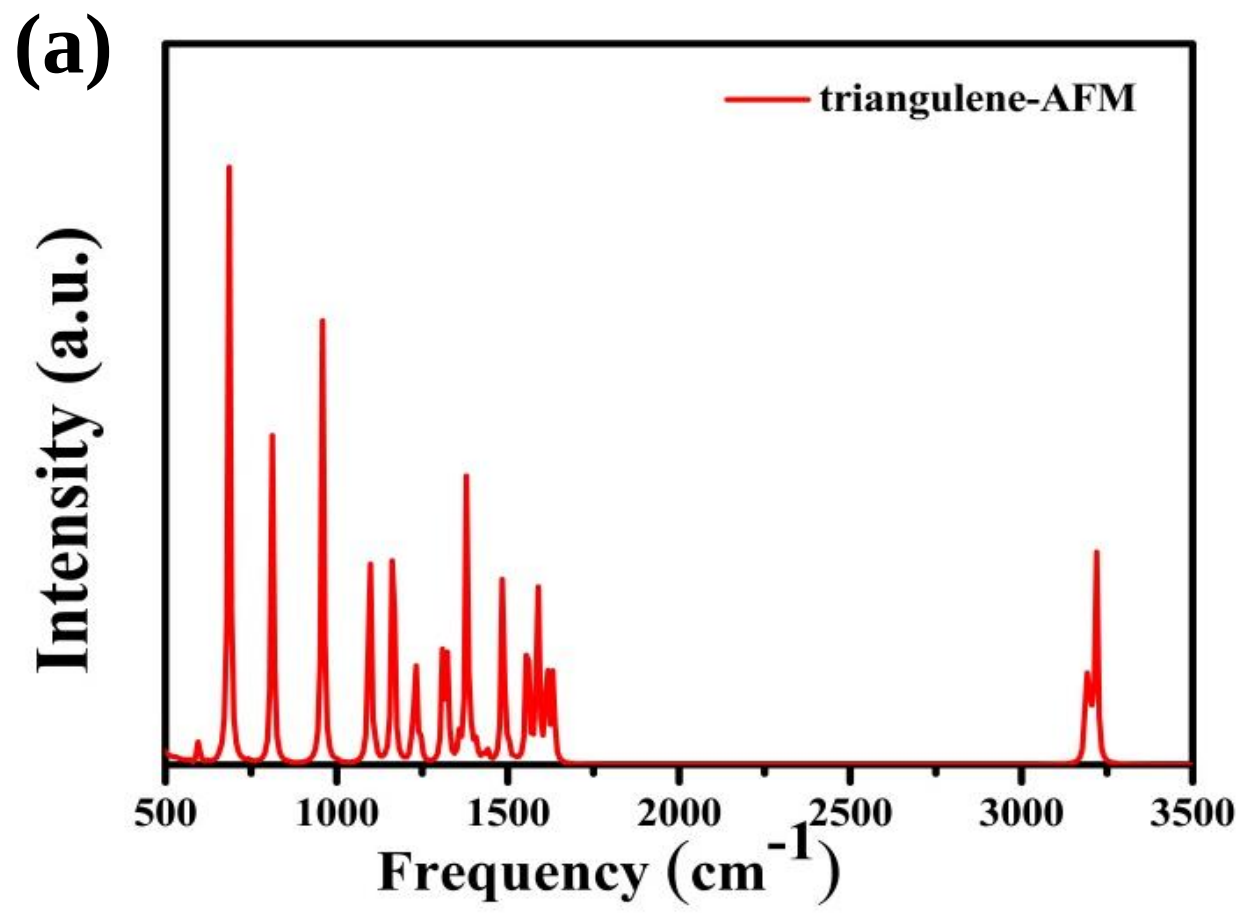

**Fig. 3**

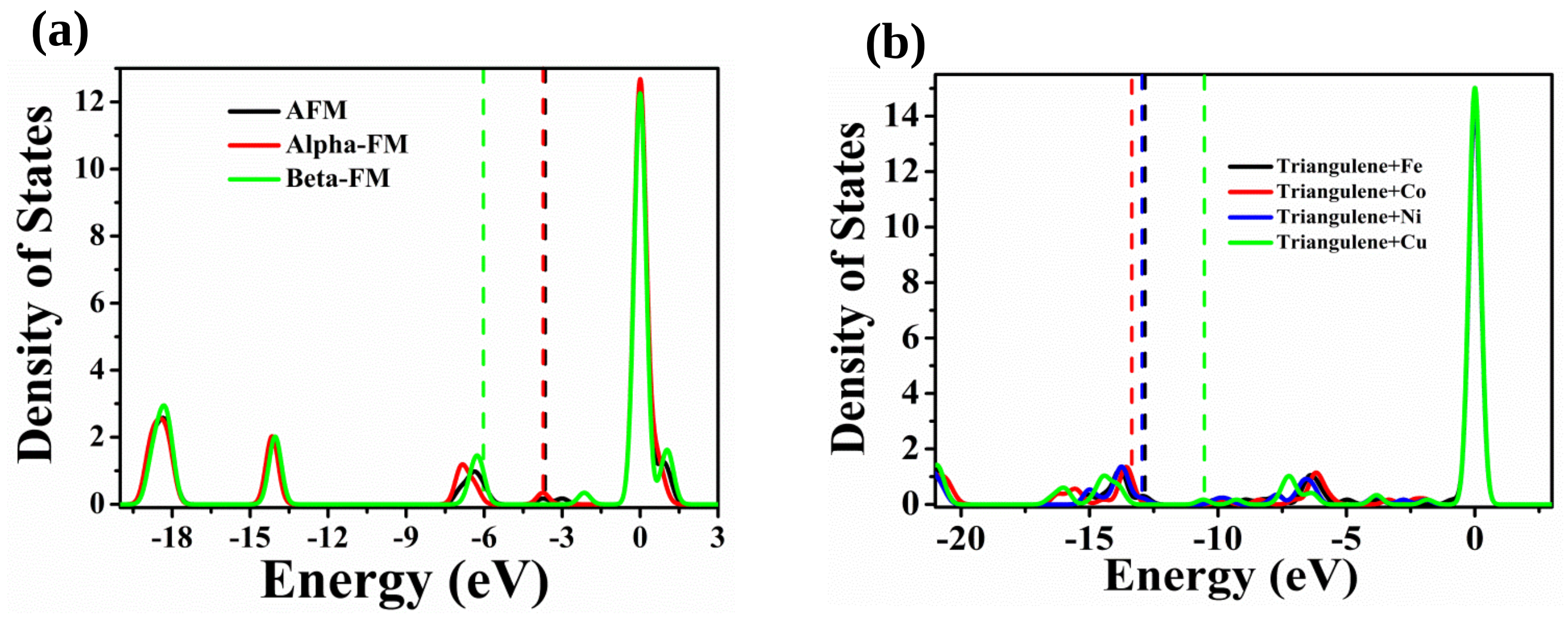

**Fig. 4**

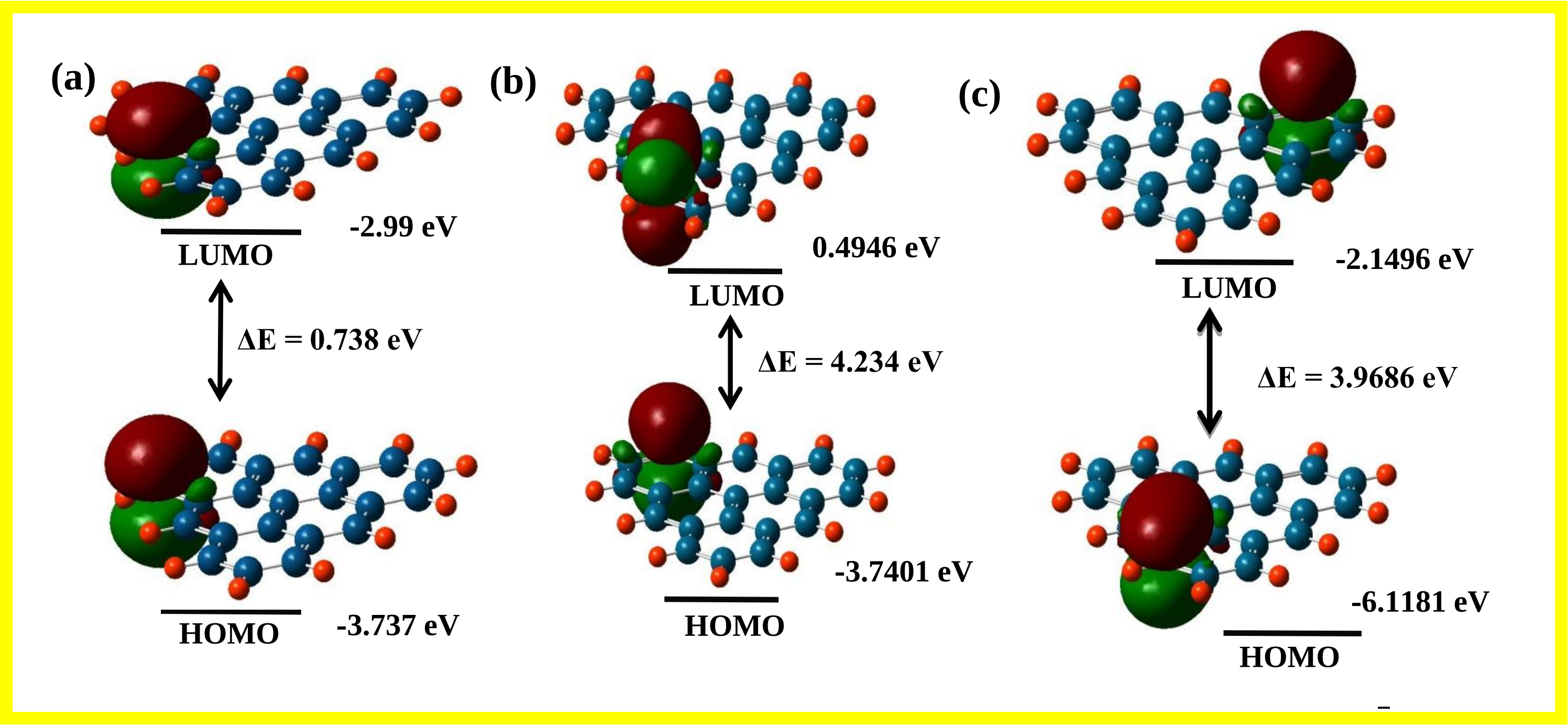

Fig. 5

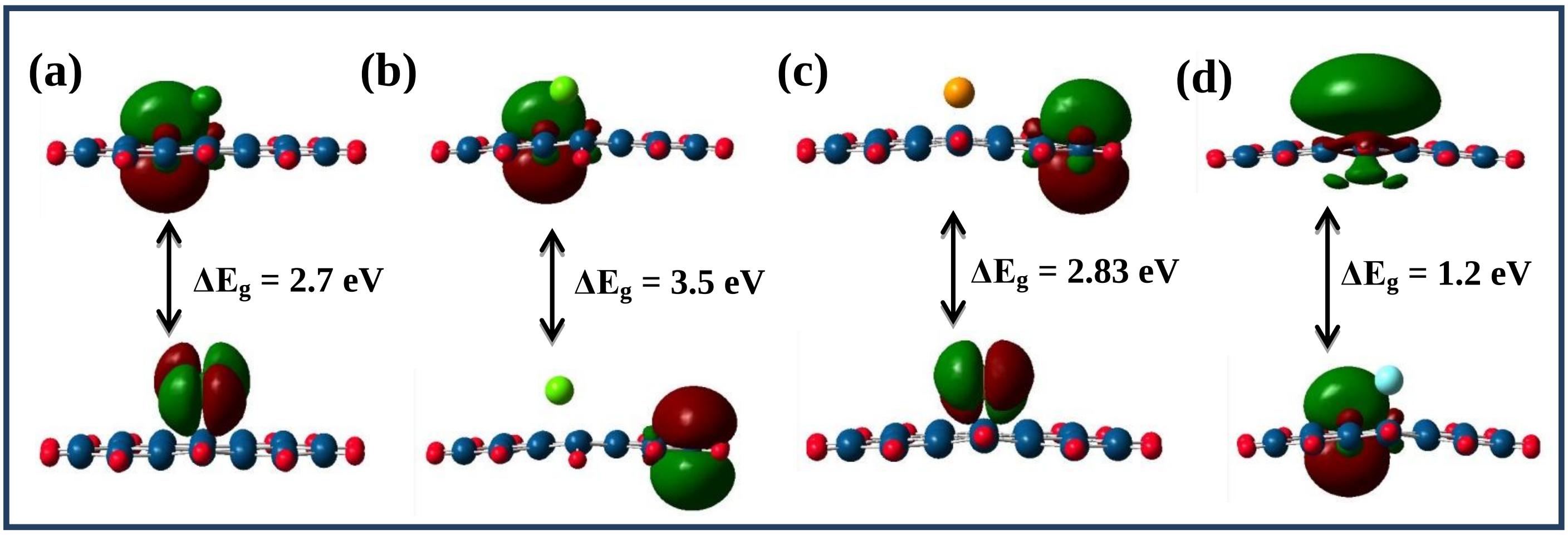
(a)
(b)
(c)
(d)
$\Delta E_g$ = 2.7 eV
$\Delta E_g$ = 3.5 eV
$\Delta E_g$ = 2.83 eV
$\Delta E_g$ = 1.2 eV

Fig. 6

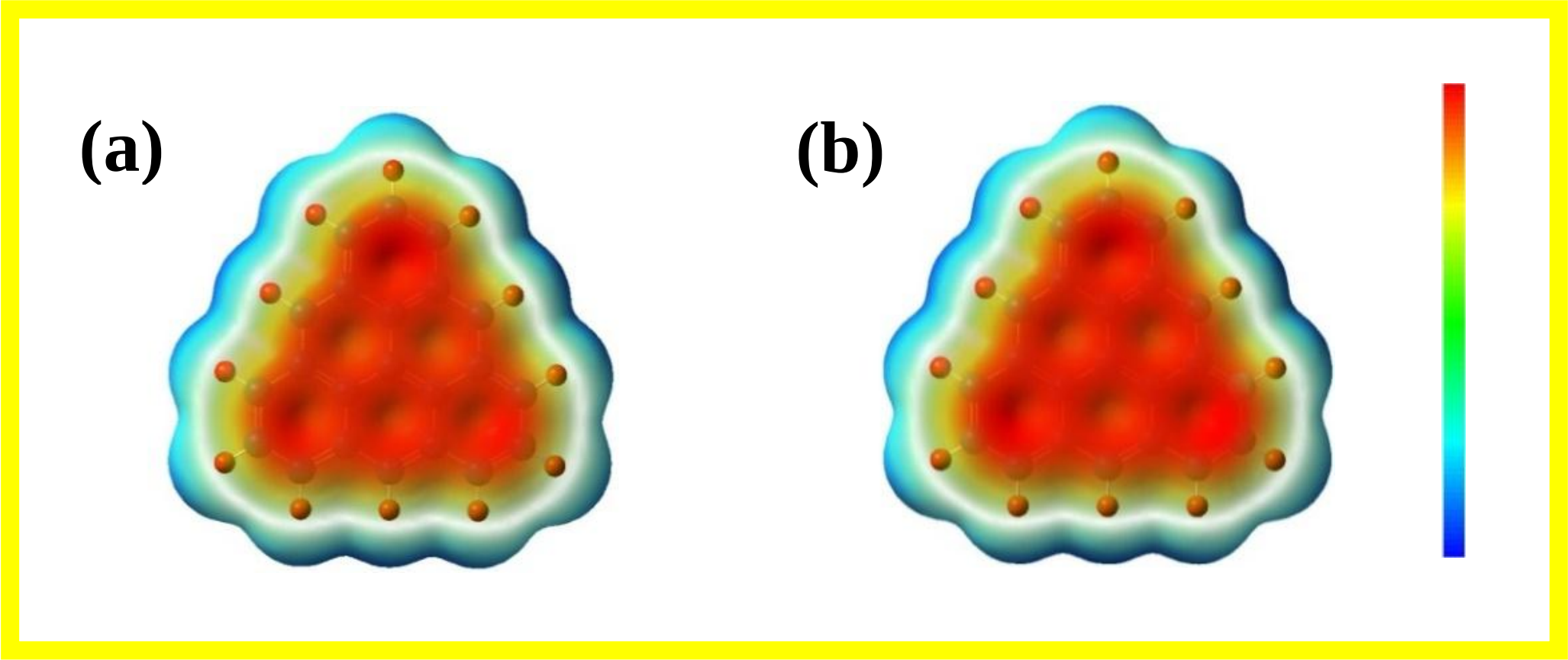
(a)
(b)

**Fig. 7**

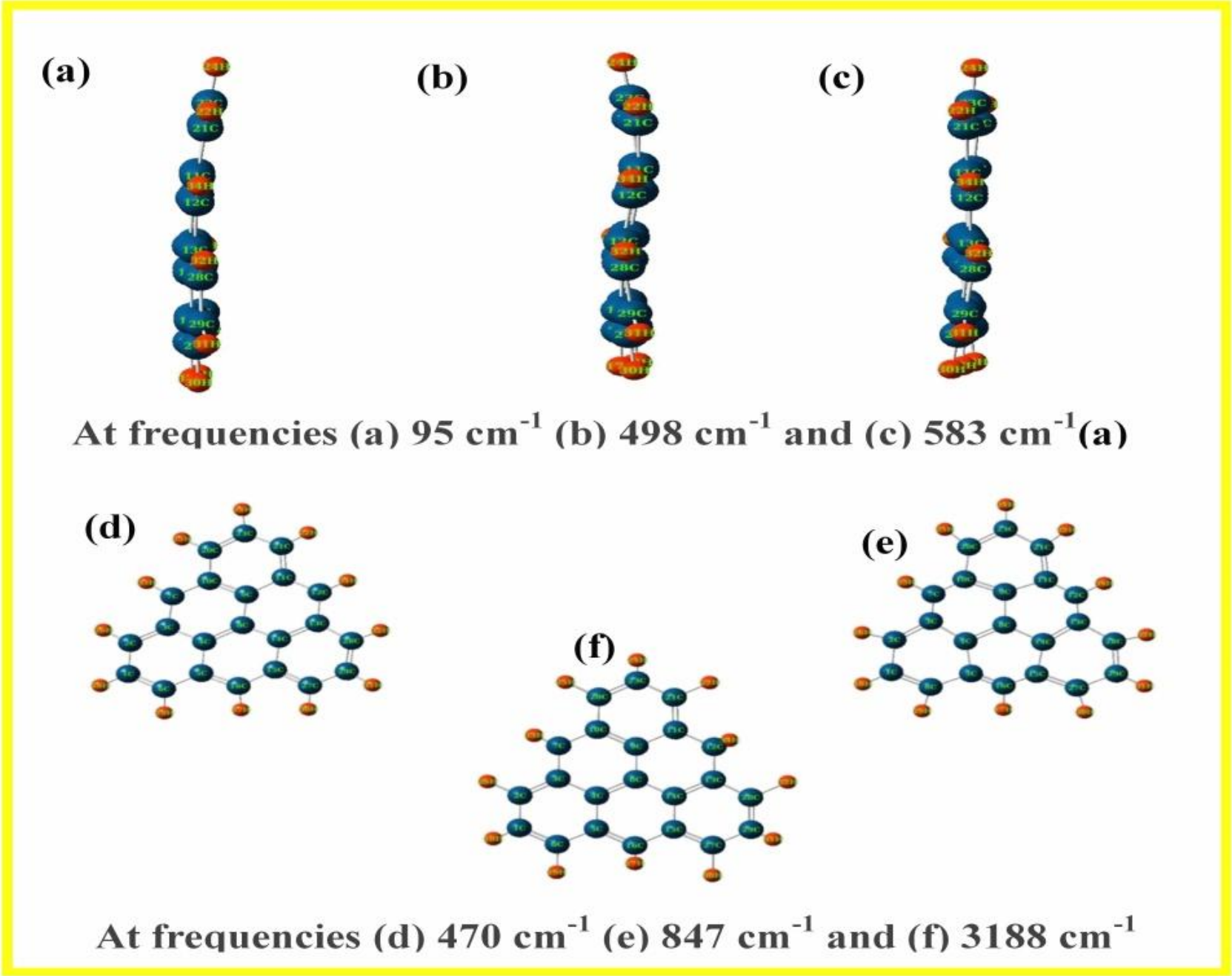
(a)
(b)
(c)
At frequencies (a) 95 cm-1 (b) 498 cm-1 and (c) 583 cm-1(a)
(d)
(e)
(f)
At frequencies (d) 470 cm-1 (e) 847 cm-1 and (f) 3188 cm-1

**Fig. 8**